\documentclass[preprint,12pt]{elsarticle}

\usepackage{amssymb}
\usepackage{amsthm,amsmath}

\journal{Physics Letters A}

\newcommand{\R}{\mathbb{R}}

\newcommand{\D}{\mathrm{d}}

\newcounter{claim}
\newtheorem{theorem}[claim]{Theorem}

\newtheorem{lemma}[claim]{Lemma}

\begin{document}

\begin{frontmatter}

%% Title, authors and addresses

%% use the tnoteref command within \title for footnotes;
%% use the tnotetext command for theassociated footnote;
%% use the fnref command within \author or \address for footnotes;
%% use the fntext command for theassociated footnote;
%% use the corref command within \author for corresponding author footnotes;
%% use the cortext command for theassociated footnote;
%% use the ead command for the email address,
%% and the form \ead[url] for the home page:
%% \title{Title\tnoteref{label1}}
%% \tnotetext[label1]{}
%% \author{Name\corref{cor1}\fnref{label2}}
%% \ead{email address}
%% \ead[url]{home page}
%% \fntext[label2]{}
%% \cortext[cor1]{}
%% \address{Address\fnref{label3}}
%% \fntext[label3]{}

\title{Spectral asymptotics of a strong $\delta'$ interaction supported by
a surface}

%% use optional labels to link authors explicitly to addresses:
%% \author[label1,label2]{}
%% \address[label1]{}
%% \address[label2]{}

\author{Pavel Exner}
\address{Doppler Institute for Mathematical Physics and Applied
Mathematics, \\ Czech Technical University in Prague,
B\v{r}ehov\'{a} 7, 11519 Prague, \\ and  Nuclear Physics Institute
ASCR, 25068 \v{R}e\v{z} near Prague, Czech republic}
\ead{exner@ujf.cas.cz}

%    Information for second author
\author{Michal Jex}
\address{Doppler Institute for Mathematical Physics and Applied
Mathematics, \\ Czech Technical University in Prague,
B\v{r}ehov\'{a} 7, 11519 Prague, \\ and Department of Physics,
Faculty of Nuclear Sciences and Physical Engineering, Czech
Technical University in Prague, B\v{r}ehov\'{a} 7, 11519 Prague, ,
Czech republic} \ead{jexmicha@fjfi.cvut.cz}

\begin{abstract}
We derive asymptotic expansion for the spectrum of Hamiltonians with a strong attractive $\delta'$ interaction supported by a smooth surface in $\R^3$, either infinite and asymptotically planar, or compact and closed. Its second term is found to be determined by a Schr\"odinger type operator with an effective potential expressed in terms of the interaction support curvatures.

\end{abstract}

\begin{keyword}
$\delta'$ surface interaction, strong coupling expansion 

%% PACS codes here, in the form: \PACS code \sep code

%% MSC codes here, in the form: \MSC code \sep code
%% or \MSC[2008] code \sep code (2000 is the default)

\end{keyword}

\end{frontmatter}

%% \linenumbers

%% main text
%%%%%%%%%%%%%%%%%%%%%%%%%%%%%%%%%%%%%%%%%%%%%%%%%%%%%%%%%%
%%  INTRODUCTION                                        %%
%%%%%%%%%%%%%%%%%%%%%%%%%%%%%%%%%%%%%%%%%%%%%%%%%%%%%%%%%%

\section{Introduction}

\noindent  Quantum mechanics of particles confined to curves,
graphs, tubes, surfaces, layers, and other geometrically nontrivial
objects is a rich and inspirative subject. On one hand it is useful
physically, in particular, to describe various nanostructures, and
at the same time it offers numerous interesting mathematical
problems. Models of ``leaky'' structures \cite{Ex} in which the
confinement is realized by an attractive potential have the
advantage that they take quantum tunneling into account. The
potential is often taken singular, of the $\delta$ type, because it
is easier to handle \cite{BLL}.

Very recently also more singular couplings of the $\delta'$ type attracted attention. The corresponding Hamiltonians can be formally written as
 % ------------- %
\begin{equation} \label{formHam}
H_\beta =-\Delta-\beta^{-1}\delta'(\cdot-\Gamma)\,,
\end{equation}
 % ------------- %
where $\Gamma$ is a smooth surface supporting the interaction. A proper definition which employs the standard $\delta'$ concept \cite{AGHH} will be given below, here we only note that a strong $\delta'$ interaction corresponds to \emph{small} values of the parameter~$\beta$. We also note that investigation of such $\delta'$ interactions is not just a mathematical exercise. Due to a seminal idea of Cheon and Shigehara \cite{CS} made rigorous in \cite{AN, ENZ} they can be approximated by a scaled ``tripple-layer'' potential combination. The possibility of forming such systems with barriers which become more opaque as the energy increases is no doubt physically attractive.

The subject of this letter is the strong coupling asymptotics of
bound states of operators \eqref{formHam} with an attractive
$\delta'$ interaction supported by a finite or infinite surface in
$\mathbb R^3$. The analogous problem for $\delta$ interaction
supported by infinite surface was solved in \cite{EK}. As in this
case, we are going to show that the asymptotics is determined by the
geometry of $\Gamma$. As a byproduct, we will demonstrate the
existence of bound states for sufficiently small $\beta$ for
non-planar infinite surfaces which are asymptotically planar, in a
way alternative to the argument proposed recently in \cite{BEL}.

%%%%%%%%%%%%%%%%%%%%%%%%%%%%%%%%%%%%%%%%%%%%%%%%%%%%%%%%%%
%%  Hamiltonian                                         %%
%%%%%%%%%%%%%%%%%%%%%%%%%%%%%%%%%%%%%%%%%%%%%%%%%%%%%%%%%%

\section{The Hamiltonian}

\noindent The first thing to do is to define properly the operator~\eqref{formHam}. It acts, of course, as Laplacian outside of the surface $\Gamma$
 % ------------- %
\begin{equation*}
(H_\beta\psi)(x)=-(\Delta\psi)(x)
\end{equation*}
 % ------------- %
for $x\in\mathbb{R}^3\setminus\Gamma$ and the interaction will be expressed through suitable boundary conditions on the surface which, in accord with \cite{AGHH}, would include continuity of the normal derivative together with a jump of the function value. Specifically, the domain of the operator will be
 % ------------- %
\begin{eqnarray*}
\lefteqn{\mathcal{D}(H_\beta)=\{\psi\in H^2(\mathbb{R}^3\setminus\Gamma)
\mid \partial_{n_\Gamma}\psi(x)=\partial_{-n_\Gamma}\psi(x)=:\psi'(x)|_{\Gamma},}
\\ &&  \hspace{7em} -\beta\psi'(x)|_{\Gamma}=\psi(x)|_{\partial_+\Gamma}
-\psi(x)|_{\partial_-\Gamma}\}\,,
\end{eqnarray*}
 % ------------- %
where $n_\Gamma$ is the normal to $\Gamma$ and $\psi(x)|_{\partial_\pm\Gamma}$ are the appropriate traces of the function $\psi$; all these quantities exist in view of the Sobolev embedding theorem. Being interested in the attractive $\delta'$ interactions, we choose the above form of boundary conditions with $\beta>0$. Another way to define the operator $H_\beta$ is by the means of the associated quadratic form as discussed in \cite{BLL}. The form value for a function $\psi\in H^1(\mathbb{R}^3\setminus\Gamma)$ is given by
 % ------------- %
\begin{equation}
\label{form} h_\beta[\psi]= \|\nabla \psi\|^2
-\beta^{-1}\|\psi(x)|_{\partial_+\Gamma}
-\psi(x)|_{\partial_-\Gamma}\|^2_{L^2(\Gamma)}\,.
\end{equation}
 % ------------- %

As indicated we are interested in the spectrum of $H_\beta$ in the
strong-coupling regime, $\beta\rightarrow0_+$, for two kinds of
surfaces $\Gamma$. The first is an infinite surface of which we
assume that:
 % ------------- %
\begin{description}
\setlength{\itemsep}{-5pt}
 % ------------- %
\item{(a1)} $\Gamma$ is $C^4$ smooth and allows a global normal parametrization with uniformly bounded elliptic tensor,
 % ------------- %
\item{(a2)} $\Gamma$ has no ``near self-intersections'', i.e. there exists its symmetric layer neighborhood of a finite thickness which does not intersect with itself,
 % ------------- %
\item{(a3)} $\Gamma$ is asymptotically planar in the sense that its curvatures vanish as the geodetic distance from a fixed point tends to infinity, and finally
 % ------------- %
\item{(a4)} trivial case is excluded, $\Gamma$ is not a plane.
\end{description}
 % ------------- %
In fact, the assumption (a1) can be weakened in a way similar to \cite{CEK}, however, for the sake of simplicity we stick to the existence of a global normal parametrization. The second class to consider are finite surfaces. The compactness makes the assumptions simpler in this case, on the other hand, we have to require additionally absence of a boundary:
 % ------------- %
\begin{description}
%\setlength{\itemsep}{-5pt}
 % ------------- %
\item{(b)} $\Gamma$ is a closed $C^4$ smooth surface of a finite genus.
 % ------------- %
\end{description}
 % ------------- %
In this case no global parametrization exists, of course, but the geometry of $\Gamma$ can be described by an atlas of maps representing normal parametrizations with a uniformly bounded elliptic tensor.

%%%%%%%%%%%%%%%%%%%%%%%%%%%%%%%%%%%%%%%%%%%%%%%%%%%%%%%%%%
%%  Preliminaries                                       %%
%%%%%%%%%%%%%%%%%%%%%%%%%%%%%%%%%%%%%%%%%%%%%%%%%%%%%%%%%%

\section{Geometric preliminaries}

\noindent Let us collect now some needed facts about the geometry of
the surface and its neighborhoods; for a more complete information
we refer, e.g., to \cite{F}. We consider infinite surfaces first and
we introduce normal coordinates on $\Gamma$ starting from a local
exponential map $\gamma:T_o\Gamma\rightarrow U_o$ with the origin
$o\in\Gamma$ to the neighborhood $U_o$ of the point $o$; the
coordinates $s$ are given by
 %----------------%
\begin{equation} \label{normcoor}
s=(s_1,s_2)\rightarrow\exp_o\big(\sum\limits_{i}s_ie_i(o)\big)
\end{equation}
 %----------------%
where $\{e_1(o),e_2(o)\}$ is an orthonormal basis of $T_o\Gamma$. By
assumption (a1) one can find a point $o\in\Gamma$  such that the map
\eqref{normcoor} can be extended to a global diffeomorphism from
$T_o\Gamma\simeq\mathbb R^2$ to $\Gamma$.

Using these coordinates, we express components of the surface metric tensor $g_{\mu\nu}$ as $g_{\mu\nu}=\gamma_{,\mu}\cdot\gamma_{,\nu}$ and denote   $g^{\mu\nu} =(g_{\mu\nu})^{-1}$. The invariant surface element is denoted as $\D\Gamma=g^{\frac{1}{2}}\D^2s$ where $g:=\det g_{\mu\nu}$. The unit normal $n(s)$ is defined as the cross product of the linearly independent tangent vectors $\gamma_{,\mu}$, i.e. $n(s)=\frac{\gamma_{,1} \times\gamma_{,2}} {|\gamma_{,1}\times\gamma_{,2}|}$. The Gauss curvature $K$ and mean curvature $M$ can be calculated by means of the Weingarten tensor  $h_\mu^\nu:= -n_{,\mu}\cdot\gamma_{,\sigma}g^{\sigma\nu}$,
 %----------------%
\begin{equation*}
K=\det h_\mu^\nu=k_1k_2\,,\quad
M=\frac{1}{2}\mathrm{Tr\,}h_\mu^\nu=\frac{1}{2}(k_1+k_2)\,.
\end{equation*}
 %----------------%
We recall that the eigenvalues of $h_\mu^\nu$ are the principal curvatures $k_{1,2}$ and that the identity $K-M^2=-\frac 1 4 (k_1-k_2)^2$ holds.

We also need neighborhoods of the surface $\Gamma$. A layer $\Omega_d$ of halfwidth $d>0$ will be defined as the image of $D_{d}:=\{(s,u):s\in\mathbb R^2,u\in(-d,d)\}$ by the map
 %----------------%
\begin{equation}
\label{curvilin}
\mathcal L:\:D_{d}\ni q\equiv(s,u)\rightarrow \gamma(s)+un(s)
\end{equation}
 %----------------%
This definition provides at the same time a parametrization of $\Omega_d$, and the assumption (a2) can be rephrased as
 % ------------- %
\begin{description}
%\setlength{\itemsep}{-5pt}
 % ------------- %
\item{(a2)} there is a $d_0>0$ such that the map \eqref{curvilin} is injective for any $d<d_0$.
 % ------------- %
\end{description}
 % ------------- %
Moreover, in view of (a1) such an $\mathcal{L}$ is a diffeomorphism, which will be crucial for the considerations to follow. The layer $\Omega_d$ can be regarded as a manifold with a boundary and characterized by the metric tensor which can be expressed in the
parametrization (\ref{curvilin}) as
 %----------------%
\begin{equation*}
G_{ij}=\left(
\begin{array}{cc}
  (G_{\mu\nu}) & 0 \\
  0 & 1 \\
\end{array}\right)\,,
\end{equation*}
 %----------------%
where $G_{\mu\nu}=(\delta_\mu^\sigma-uh_\mu^\sigma)(\delta_\rho^\sigma-uh_\rho^\sigma)g_{\rho\nu}$. We use here the convention in which the Latin indices run through $1,2,3$, numbering the coordinates $(s_1,s_2,u)$ in $\Omega_d$, and the Greek ones through $1,2$. The volume element of the manifold $\Omega_d$ can be written in the form $\D\Omega_d:=\sqrt{G}\,\D^2s\,\D u$ with
 %----------------%
\begin{equation*}
G:=\det G_{ij}=g[(1-uk_1)(1-uk_2)]^2=g(1-2Mu+Ku^2)^2\,;
\end{equation*}
 %----------------%
with the future purpose in mind we introduce a shorthand for the last factor,  $\xi(s,u):=1-2M(s)u+K(s)u^2$. The curvatures also allow us to express more explicitly the next assumption:
 % ------------- %
\begin{description}
\setlength{\itemsep}{-5pt}
 % ------------- %
\item{(a3)} $K,M\rightarrow0$ as $|s|:= \sqrt{s_1^2+s_2^2}\rightarrow\infty$.
 % ------------- %
\end{description}
 % ------------- %

Recall next a few useful estimates made possible by the assumption (a3), cf.~\cite{DEK}. In combination with (a1) and (a2) it implies that the principal curvatures $k_1$ and $k_2$ are uniformly bounded. We set
 %----------------%
\begin{equation*}
\rho:=(\max\{\|k_1\|_{\infty},\|k_2\|_{\infty}\})^{-1}\,;
\end{equation*}
 %----------------%
note that $\rho>d_0$ holds for the critical halfwidth of assumption (a2). It can be checked easily that for a given $d<\rho$ the following inequalities are satisfied in the layer neighborhood $\Omega_d$ of $\Gamma$,
 %----------------%
\begin{equation}
\label{xi} C_-(d)\leq\xi\leq C_+(d)\,,
\end{equation}
 %----------------%
where $C_\pm:=(1\pm d\rho^{-1})^2$, and this in turn implies
 %----------------%
\begin{equation}
\label{Gmunu} C_-(d)g_{\mu\nu}\leq G_{\mu\nu}\leq C_+(d)g_{\mu\nu}\,.
\end{equation}
 %----------------%
Since the metric tensor $g_{\mu\nu}$ uniformly elliptic by assumption, we also have
 %----------------%
\begin{equation}
\label{gmunu} c_-\delta_{\mu\nu}\leq g_{\mu\nu}\leq
c_+(d)\delta_{\mu\nu}
\end{equation}
 %----------------%
as a matrix inequality for some positive constants $c_\pm$.

Let us briefly describe modifications needed if we pass to closed
surfaces. As we have indicated a global parametrization is replaced now by a finite atlas $\mathcal A$ of maps; in each part $\mathcal M_i$ we
introduce normal coordinates and define layer neighborhoods by the
maps $\hat{\mathcal M_i}$ on $D_{i,d}:=\{(s,u):\,
s\in\mathrm{dom}\mathcal M_i,u\in(-d,d)\}$ with a given $d>0$,
  %----------------%
\begin{equation}
\label{curvilin2}
 \hat {\mathcal M_i}:\mathrm D_{i,d}\ni q\equiv(s,u)\rightarrow
\gamma_i(s)+un(s)
\end{equation}
 %----------------%
In view of assumption (b) there is a critical $d_0>0$ such that every map $\hat{\mathcal M_i}:D_{i,d}\rightarrow\Omega_d$ from $\mathcal A$ is injective provided $d<d_0$ and a diffeomorphism. Furthermore, $\hat{\mathcal M_i}(s_i,u_i)=\hat{\mathcal M_j}(s_j,u_j)$ implies $\mathcal M_i(s_i)=\mathcal M_j(s_j)$. The above estimates of the metric tensor remains valid also for compact $\Gamma$.

%%%%%%%%%%%%%%%%%%%%%%%%%%%%%%%%%%%%%%%%%%%%%%%%%%%%%%%%%%
%%  Results                                             %%
%%%%%%%%%%%%%%%%%%%%%%%%%%%%%%%%%%%%%%%%%%%%%%%%%%%%%%%%%%

\section{The results}

\noindent As in the case of a $\delta$ interaction supported by a surface, the asymptotics is determined by the geometry of $\Gamma$. To state the results, we introduce the following comparison operator,
 % ------------- %
\begin{equation}\label{comparison}
S=-\Delta_{\Gamma}-\frac{1}{4}(k_1-k_2)^2 =-\Delta_{\Gamma}+K-M^2\,,
\end{equation}
 % ------------- %
where $\Delta_\Gamma$ is the Laplace-Bertrami operator on the surface
$\Gamma$ and $k_{1,2}$ are the principal curvatures of  $\Gamma$. The spectrum of $S$ is purely discrete if $\Gamma$ is
compact. In the noncompact case the potential vanishes at infinity
and has negative values unless $\Gamma$ is a plane which is,
however, excluded by assumption (a4). Consequently,
$\sigma_\mathrm{ess}(S)= [0,\infty)$ and the discrete spectrum is
nonempty. We denote the eigenvalues of $S$, arranged in the
ascending order with the multiplicity taken into account, as
$\mu_j$.

First we inspect the essential spectrum in the strong-coupling regime:

 % ------------- %
\begin{theorem} \label{thm1}
Let an infinite surface $\Gamma$ satisfy assumptions (a1)--(a4),
then $\sigma_\mathrm{ess}(H_\beta) \subseteq
[\epsilon(\beta),\infty)$, where $\epsilon(\beta)\to
-\frac{4}{\beta^{2}}$ holds as $\beta\to 0_+$.
\end{theorem}
 % ------------- %

\noindent We note that in case of a compact $\Gamma$ we have $\sigma_\mathrm{ess}(H_\beta)=[0,\infty)$; a proof can be found in \cite{BEL}. The next two theorems describe the asymptotics of the negative point spectrum of $H_\beta$.

 % ------------- %
\begin{theorem} \label{thm2}
Let an infinite surface $\Gamma$ satisfy assumptions (a1)--(a4),
then $H_\beta$ has at least one isolated eigenvalue below the
threshold of the essential spectrum for all sufficiently small
$\beta>0$, and the $j$-th eigenvalue behaves in the limit $\beta\to
0_+$ as
 %----------------%
\begin{equation*}
\lambda_j=-\frac{4}{\beta^2}+\mu_j+\mathcal O(-\beta\ln\beta)\,.
\end{equation*}
 %----------------%
\end{theorem}
 % ------------- %
\begin{theorem} \label{thm3}
Let a compact surface $\Gamma$ satisfy assumption (b), then
$H_\beta$ has at least one isolated eigenvalue below the threshold
of the essential spectrum for all $\beta>0$, and the $j$-th
eigenvalue behaves in the limit $\beta\to 0_+$ as
 %----------------%
\begin{equation*}
\lambda_j=-\frac{4}{\beta^2}+\mu_j+\mathcal O(-\beta\ln\beta)\,.
\end{equation*}
 %----------------%
\end{theorem}

%%%%%%%%%%%%%%%%%%%%%%%%%%%%%%%%%%%%%%%%%%%%%%%%%%%%%%%%%%
%%  Bracketing                                          %%
%%%%%%%%%%%%%%%%%%%%%%%%%%%%%%%%%%%%%%%%%%%%%%%%%%%%%%%%%%

\section{Bracketing estimates}

\noindent The basic idea is analogous to the one used in \cite{EK}, namely to estimate the operator $H_\beta$ from above and below, in a tight enough manner, by suitable operators for which we are able to calculate the spectrum directly. The starting point for such estimates is the bracketing trick, that is, imposing additional Dirichlet/Neumann conditions at the boundary of the neighborhood $\Omega_d$ of the surface $\Gamma$. We introduce quadratic forms $h_\beta^+$ and $h_\beta^-$, both of them given by the formula
 % ------------- %
\begin{equation*}
\|\nabla \psi\|^2_{L^2(\Omega_d)}
-\beta^{-1}\int_{\Gamma}|\psi(s,0_+)-\psi(s,0_-)|^2\,\D \Gamma\,.
\end{equation*}
 % ------------- %
with the domains $\mathcal D(h_\beta^+)=\tilde{H}^1_0(\Omega_d\setminus\Gamma)$ and $\mathcal D(h_\beta^-)=\tilde{H}^1(\Omega_d\setminus\Gamma)$, respectively. We denote the self-adjoined operators associated with these forms as $H^\pm_\beta$. By the standard bracketing argument we get
 % ------------- %
\begin{equation}
\label{nerovnost} -\Delta_{\mathbb R^3\setminus\Omega_d}^N\oplus
H_\beta^-\leq H_\beta\leq-\Delta_{\mathbb
R^3\setminus\Omega_d}^D\oplus H_\beta^+\,,
\end{equation}
 % ------------- %
where $-\Delta_{\mathbb R^3\setminus\Omega_d}^{D,N}$ is the
Dirichlet Laplacian and Neumann Laplacian respectively on the
set${\mathbb R^3\setminus\Omega_d}$. The operators $-\Delta_{\mathbb
R^3\setminus\Omega_d}^{D,N}$ are positive, hence all the information
about the negative spectrum is encoded in the
operators~$H^\pm_\beta$.

The next step is to transform the operators $H^\pm_\beta$ into the curvilinear coordinates $(s,u)$. This is done by means of the unitary transformation
 % ------------- %
\begin{equation*}
U\psi=\psi\circ\mathcal L: L^2(\Omega_d)\rightarrow L^2(D_d,\D
\Omega).
\end{equation*}
 % ------------- %
By $(\cdot,\cdot)_G$ we denote the scalar product in $L^2(D_d,\D \Omega)$. The operators $UH^\pm_\beta U^{-1}$ acting on this space are associated with the forms
 % ------------- %
\begin{equation*}
h_\beta^\pm(U^{-1}\psi) = (\partial_i\psi,G^{ij}\partial_j\psi)_G
-\beta^{-1}\int_{\Gamma}|\psi(s,0_+)-\psi(s,0_-)|^2\,\D\Gamma
\end{equation*}
 % ------------- %
having the domains $\tilde{H}^1_0(D_d\setminus\Gamma,\D \Omega)$ and $\tilde{H}^1(D_d\setminus\Gamma,\D \Omega)$, respectively. Next we employ another unitary transformation, inspired by \cite{DEK}, with the aim to get rid of the transverse coordinate dependence, i.e. switch from the metric $\D\Omega$ to $\D\Gamma\,\D u$ by
 % ------------- %
\begin{equation*}
\tilde{U}\psi=\xi^{\frac1 2}\psi: L^2(D_d,\D\Omega)\rightarrow
L^2(D_d,\D \Gamma\D u).
\end{equation*}
 % ------------- %
Similarly as before, we denote the scalar product in $L^2(D_d,\D\Gamma\D u)$ as $(\cdot,\cdot)_g$ and consider the operators
 % ------------- %
\begin{equation*}
F_{\beta}^{\pm}:=\tilde UUH^\pm_\beta U^{-1}\tilde U^{-1}
\end{equation*}
 % ------------- %
which act in $L^2(D_d,\D \Gamma\D u)$. The quadratic forms $\zeta_\beta^\pm$ associated with $F_{\beta}^{\pm}$ can be calculated as $h_\beta^\pm(\tilde U^{-1}U^{-1}\psi)$ with the result
 % ------------- %
\begin{eqnarray*}
&& \hspace{-2em} \zeta_\beta^+[\psi]
=(\partial_\mu\psi,G^{\mu\nu}\partial_\nu\psi)_g+(\psi,(V_1+V_2)\psi)_g
+\|\partial_3\psi\|_g\\ &&
-\beta^{-1}\int_{\Gamma}|\psi(s,0_+)-\psi(s,0_-)|^2\,\D
\Gamma-\int_{\Gamma}M(|\psi(s,0_+)|^2-|\psi(s,0_-)|^2)\,\D \Gamma \\
&& \hspace{-2em} \zeta_\beta^-[\psi] = \zeta_\beta^+[\psi]
+\int_\Gamma\varsigma(s,d)|\psi(s,d)|^2\, \D \Gamma
-\int_\Gamma\varsigma(s,-d)|\psi(s,-d)|^2\, \D \Gamma\,.
\end{eqnarray*}
 % ------------- %
where $\varsigma=\frac{M-Ku}{\xi}$, the two curvature-induced potentials are
 % ------------- %
\begin{equation*}
V_1=g^{-\frac 1 2}(g^{\frac 1
2}G^{\mu\nu}J_{,\mu})_{,\nu}+J_{,\mu}G^{\mu\nu}J_{,\nu}\,, \quad
V_2=\frac{K-M^2}{\xi^2}
\end{equation*}
 % ------------- %
with $J=\frac{\ln \xi}{2}$. The corresponding form domains are $\tilde{H}^1_0(D_d\setminus\Gamma,\D \Gamma\D u)$ and $\tilde{H}^1(D_d\setminus\Gamma,\D \Gamma\D u)$, respectively.

%%%%%%%%%%%%%%%%%%%%%%%%%%%%%%%%%%%%%%%%%%%%%%%%%%%%%%%%%%
%%  First proof                                         %%
%%%%%%%%%%%%%%%%%%%%%%%%%%%%%%%%%%%%%%%%%%%%%%%%%%%%%%%%%%

\section{Proof of Theorem~\ref{thm1}}

\noindent In the excluded case when $\Gamma$ is a plane, the
spectrum is easily found by separation of variables which gives
$\sigma(H_\beta) = \sigma_\mathrm{ess}(H_\beta) =
\big[-\frac{4}{\beta^2},\infty\big)$. We want to show that under the
assumption (a3) the essential spectrum does not change, at least
asymptotically. We employ an estimate which follows from
Lemma~\ref{trans} that we will prove below, namely
 % ------------- %
\begin{equation}
\label{ner2}\int\limits_{-d}^d \left|\frac{\D f}{\D u}\right|^2\D
u-\beta^{-1}|f(0_+)-f(0_-)|^2\geq
\left(-\frac{4}{\beta^2}-\frac{16}{\beta^2}\exp\left(-\frac{4d}{\beta}\right)\right)\|f\|_{L^2(-d,d)}
\end{equation}
 % ------------- %
As we shall see the inequality holds for sufficiently small $\beta$ and $\frac{d}{\beta}>2$. The inclusion $\sigma_\mathrm{ess}(H_\beta) \subseteq
[\epsilon(\beta),\infty)$ is equivalent to
 % ------------- %
\begin{equation*}
\inf\sigma_\mathrm{ess}(H_\beta) \geq \epsilon(\beta)
\end{equation*}
 % ------------- %
which will be satisfied if $\inf\sigma_\mathrm{ess}(H^-_\beta) \geq \epsilon(\beta)$ for $H^-_\beta$ acting in $L^2(\Omega_d)$ for $d<g_0<\rho$. This is obvious from inequalities (\ref{nerovnost}) and the positivity of $-\Delta_{\mathbb R^3\setminus\Omega_d}^N$. In the next step we divide  the surface $\Gamma$ into two parts, namely $\Gamma_\tau^\mathrm{int}:=\{s\in\Gamma|\,r(s)<\tau\}$ and $\Gamma_\tau^\mathrm{ext} :=\Gamma\setminus \overline{\Gamma_\tau^\mathrm{int}}$. The layer neighborhoods corresponding to $\Gamma_\tau^\mathrm{int}$ and $\Gamma_\tau^\mathrm{ext}$ are $D_\tau^\mathrm{int}=\{(s,u)\in D_d|\,s\in\Gamma_\tau^\mathrm{int}\}$ and $D_\tau^\mathrm{ext}=D_d\setminus\overline{D_\tau^\mathrm{int}}$. We introduce the Neumann operators on respective neighborhoods, $H^{-,z}_{\beta,\tau}$ for $z=\mathrm{int},\mathrm{ext}$ associated with the forms
 % ------------- %
\begin{equation*}
(\partial_i\psi,G^{ij}\partial_j\psi)_G-\beta^{-1}\int_{\Gamma_\tau^z}|\psi(s,0_+)-\psi(s,0_-)|^2\,\D
\Gamma
\end{equation*}
 % ------------- %
defined on $\tilde{H}^1(D_\tau^z\setminus\Gamma,\D \Omega)$. Using once more Neumann bracketing we get $H^-_\beta\geq H^{-,\mathrm{int}}_{\beta,\tau} \oplus H^{-,\mathrm{ext}}_{\beta,\tau}$. The inner part is compact, hence the spectrum of $H^{-,\mathrm{int}}_{\beta,\tau}$ is purely discrete. Consequently, the min-max principle implies
 % ------------- %
\begin{equation*}
\inf\sigma_\mathrm{ess}(H^-_\beta) \geq
\inf\sigma_\mathrm{ess}(H^{-,\mathrm{ext}}_{\beta,\tau})\,,
\end{equation*}
 % ------------- %
and it is sufficient to check that the right-hand side cannot be smaller than $\epsilon(\beta)$. The quantities
$m_\tau^+:=\sup_{\Gamma_\tau^\mathrm{ext}}\xi$ and $m_\tau^-:=\inf_{\Gamma_\tau^\mathrm{ext}}\xi$ tend to one as
$\tau\rightarrow\infty$ in view of assumption (a3). We have the following estimate,
 % ------------- %
\begin{eqnarray*}
(\psi,H^{-,\mathrm{ext}}_{\beta,\tau}\psi)_{G}\geq
\int_{D_\tau^\mathrm{ext}}|\partial_3\psi(q)|^2\D\Omega-\beta^{-1}
\int_{\Gamma_\tau^\mathrm{ext}}|\psi(s,0_+)-\psi(s,0_-)|^2\,\D
\Gamma\\
\geq
m_\tau^-\int_{D_\tau^\mathrm{ext}}|\partial_3\psi(q)|^2\D\Gamma\D
u-\beta^{-1}\int_{\Gamma_\tau^\mathrm{ext}}|\psi(s,0_+)-\psi(s,0_-)|^2\,\D
\Gamma\\
\geq \frac{1}{\beta^2m_\tau^+m_\tau^-}\left[-\frac 1 4
-16\exp\left(-\frac{4d}{\beta}\right)\right]\int_{D_\tau^\mathrm{ext}}|\psi(q)|^2\D\Omega\,,
\end{eqnarray*}
 % ------------- %
and since $\tau$ is arbitrary, we obtain $\epsilon(\beta) \ge -\frac{4}{\beta^2}-\frac{16}{\beta^2}\exp\left(-\frac{4d}{\beta}\right)$.

%%%%%%%%%%%%%%%%%%%%%%%%%%%%%%%%%%%%%%%%%%%%%%%%%%%%%%%%%%
%%  Second proof                                         %%
%%%%%%%%%%%%%%%%%%%%%%%%%%%%%%%%%%%%%%%%%%%%%%%%%%%%%%%%%%

\section{Proof of Theorem~\ref{thm2}}

\noindent To prove the second theorem, we will need several auxiliary results. The operators $F_\beta^\pm$ are still not suitable to work with and
so we replace them with a slightly cruder bounds. First we estimate the values of the potentials $V_1$ and $V_2$. With the help of inequalities (\ref{xi})--(\ref{gmunu}) we are able to check that
 % ------------- %
\begin{equation*}
d v^-\leq V_1\leq d v^+
\end{equation*}
 % ------------- %
holds for suitable numbers $v^\pm$ and $d<d_0<\rho$. On the other hand, $V_2$ can be estimated as
 % ------------- %
\begin{equation*}
C_-^{-2}(K-M^2)\leq V_2\leq C_+^{-2}(K-M^2)\,,
\end{equation*}
 % ------------- %
where $C_\pm$ are the same as in (\ref{xi}). This allows us to replace \eqref{nerovnost} with the estimates using operators $D_\beta^\pm$,
 % ------------- %
\begin{equation}
\label{ner3}\begin{array}{c}
   D_{d,\beta}^{-} := U_{d}^-\otimes I+\int_{\Gamma}^\oplus
T_{d,\beta}^{-}(s)\,\D \Gamma \leq F_\beta^- \leq H_\beta \\[.8em]
  H_\beta\leq F_\beta^+\leq U_{d}^+\otimes I+\int_{\Gamma}^\oplus
T_{d,\beta}^{+}(s)\,\D \Gamma =: D_{d,\beta}^{+}
\end{array}
\end{equation}
 % ------------- %
where
 % ------------- %
\begin{equation*}
U_{d}^\pm=-C_\pm\Delta_\Gamma+C_\pm^{-2}(K-M^2)+v^{\pm}d
\end{equation*}
 % ------------- %
with the domain $\mathcal D(U_{d}^\pm)=L^2(\mathbb R^2,\D \Gamma)$ and the transverse part acts as
 % ------------- %
\begin{equation*}
T_{d,\beta}^{\pm}(s)\psi =-\Delta\psi
\end{equation*}
 % ------------- %
with the domains
 % ------------- %
\begin{eqnarray*}
\lefteqn{\mathcal{D}(T_{a,\beta}^{+}(s))= \Big\{f\in
H^2((-a,a)\setminus\{0\})\mid f(a)=f(-a)=0,} \\[.2em] && f'(0_-)=f'(0_+)=
-\beta^{-1}(f(0_+)-f(0_-))+M(f(0_+)+f(0_-)) \Big\}
\end{eqnarray*}
 % ------------- %
and
\begin{eqnarray*}
\lefteqn{\hspace{-1em} \mathcal{D}(T_{a,\beta}^{-}(s))= \Big\{f\in
H^2((-a,a)\setminus\{0\})\mid \mp
\frac{\|M\|_\infty+d\|K\|_\infty}{C_-}f(\pm a)=f'(\pm a),} \\ &&
f'(0_-)=f'(0_+)= -\beta^{-1}(f(0_+)-f(0_-))+M(f(0_+)+f(0_-))
\Big\}\,,
\end{eqnarray*}
 % ------------- %
respectively. The negative spectrum is described by the following result the proof of which can be found in \cite{EJ}.

 % ------------- %
\begin{lemma}\label{trans}
Each of the operators $T_{d,\beta}^{\pm}(s)$ has exactly one negative eigenvalue $t_\pm(d,\beta)$, respectively, which is independent of $s$ provided that $\frac{d}{\beta}>2$ and $\beta(\|M\|_\infty+d\|K\|_\infty)<1$. For all $\beta>0$ sufficiently small these eigenvalues satisfy the  inequalities
 %----------------%
\begin{equation*}
-\frac{4}{\beta^2}-\frac{16}{\beta^2}\exp\left(-\frac{4d}{\beta}\right)\leq
t_-(d,\beta)\leq-\frac{4}{\beta^2}\leq
t_+(d,\beta)\leq-\frac{4}{\beta^2}+\frac{16}{\beta^2}\exp\left(-\frac{4d}{\beta}\right)\,.
\end{equation*}
 %----------------%
\end{lemma}
 %----------------%

\noindent On the other hand, the spectrum of the operators $U_{d}^\pm$ has the asymptotic expansion governed by the operator $S$ which we can adopt from \cite{EK}:

 %----------------%
\begin{lemma}\label{longi}
The eigenvalues of $U_{d}^\pm$ satisfy the relations
eigenvalues
 %----------------%
\begin{equation*}
\mu_j^{\pm}(d)=\mu_j+C^\pm_jd+\mathcal O(d^2)\quad\textrm{for}\quad
d\to 0\,,
\end{equation*}
 %----------------%
where $\mu_j$ is the $j$-th eigenvalue of the operator $S$ and the constants
$C_j^\pm$ are independent on $d$.
\end{lemma}
 %----------------%

\noindent With these prerequisites we are ready to prove the second
theorem. We put $d(\beta)=-\beta\ln\beta$. Using the fact that each of the operators $T_{d,\beta}^{\pm}(s)$ has exactly one negative
eigenvalue $t_\pm(d(\beta),\beta)$ together with the explicit form
of  $D_{d,\beta}^{\pm}$ we can write their spectra as
$t_\pm(d(\beta),\beta)+\mu_j^{\pm}(d(\beta))$, where $\mu_j^{\pm}$
are the eigenvalues of the operators $U_{d}^\pm$. Using now
Lemmata~\ref{trans} and \ref{longi} we are able to rewrite this as
 %----------------%
\begin{equation*}
t_\pm(d(\beta),\beta)+\mu_j^{\pm}(d(\beta))=-\frac {4}
{\beta^{2}}+\mu_j+\mathcal O(\beta|\ln\beta|)\,,
\end{equation*}
hence the min-max principle in combination with inequalities (\ref{ner3}) conclude the argument.

%%%%%%%%%%%%%%%%%%%%%%%%%%%%%%%%%%%%%%%%%%%%%%%%%%%%%%%%%%
%%  Third proof                                         %%
%%%%%%%%%%%%%%%%%%%%%%%%%%%%%%%%%%%%%%%%%%%%%%%%%%%%%%%%%%

\section{Proof of Theorem~\ref{thm3}}

\noindent The existence of isolated eigenvalues can be checked
variationally as in \cite{BEL}. For a test function $\xi$ one
chooses characteristic function of the volume enclosed by the surface
$\Gamma$; this yields an estimate of the ground state energy from
above,
 %----------------%
\begin{equation}
\lambda_0\leq\frac{h_\beta(\xi)}{\|\xi\|^2}=\beta^{-1}\frac{S}{V}
\end{equation}
 %----------------%
where $h_\beta$ is quadratic (\ref{form}), $S$ is the  area of the surface $\Gamma$ and $V$ is the volume enclosed by $\Gamma$. The proof of the asymptotic expansion proceed with minimum modifications as for the infinite surface, hence we omit the details.

\subsection*{Acknowledgments}

\noindent The research was supported by the Czech Science Foundation  within
the project \mbox{14-06818S} and by by Grant Agency of the Czech
Technical University in Prague, grant No. SGS13/217/OHK4/3T/14.

%% If you have bibdatabase file and want bibtex to generate the
%% bibitems, please use
%%
%%  \bibliographystyle{elsarticle-harv}
%%  \bibliography{<your bibdatabase>}

%% else use the following coding to input the bibitems directly in the
%% TeX file.

\end{document}